# $H_2$: entanglement, probability density function, confined Kratzer oscillator, universal potential and Mexican hat- or bell-type potential energy curves

*G. Van Hooydonk, Ghent University, Faculty of Sciences, Ghent, Belgium*

Abstract. *We review harmonic oscillator theory for closed, stable quantum systems. The $H_2$ potential energy curve (PEC) of Mexican hat-type, calculated with a confined Kratzer oscillator, is better than the Rydberg-Klein-Rees (RKR) $H_2$ PEC. Compared with QM, the theory of chemical bonding is simplified, since a confined Kratzer oscillator gives the long sought for universal function, once called the Holy Grail of Molecular Spectroscopy. This is validated with HF, $I_2$, $N_2$ and $O_2$ PECs. We quantify the entanglement of spatially separated $H_2$ quantum states, which gives a braid view. The equal probability for $H_2$, originating either from $H_A+H_B$ or $H_B+H_A$, is quantified with a Gauss probability density function. At the Bohr scale, confined harmonic oscillators behave properly at all extremes of bound two-nucleon quantum systems and are likely to be useful also at the nuclear scale.*

**Introduction**

Since the harmonic oscillator (HO) is essential for physics [1] and chemistry [2], understanding $H_2$, the simplest, prototypical oscillator in nature, is important. While H is prototypical for atoms with simple Bohr theory, $H_2$ is prototypical for bonds but no simple oscillatory bond theory exists [2]. Inverting levels with RKR (Rydberg-Klein-Rees [3]) or IPA (inverted perturbation approximation [4]) gives the $H_2$ potential energy curve (PEC) [5]. We revisit the $H_2$ oscillator and bring in the long sought for universal potential energy function (UPEF) [6-10], once called the *Holy Grail of Molecular Spectroscopy* [11]. A new approach is needed, since QM fails on an analytical potential energy function (PEF), even for $H_2$ [5,11].

Vibrational $H_2$ levels $U_v$, nearly parabolic in quantum number v, transform in $H_2$ PEC $U_r$, nearly parabolic in r but with different curvatures and extremes. We merge all curvatures and extremes in a single PEF, using the classical ionic Kratzer-Coulomb potential [12] for $H_2$ [13]. The Kratzer $H_2$ PEC is more precise than the RKR PEC, if $2^{nd}$ order Kratzer function is upgraded to $4^{th}$ order. This gives a Mexican-hat type PEC for $H_2$ and a parameter free UPEF, although a low parameter UPEF may not even exist [6-10,11,14]. Our RKR-based solution bears on entanglement of spatially separated $H_2$ quantum states. This gives a braid effect, important for EPR-paradox and Bell inequalities [15] and quantum information theory [16], and simplifies the theory of the chemical bond [17]. This also provides with a link between the physics in a Kratzer model and probabilities (entropy). We show that the Mexican hat-type UPEC for $H_2$, HF, $I_2$, $N_2$ and $O_2$ is complementary with their bell curve for normal Gauss probability density distributions. In Section II, we review problems. Section III is on a confined Kratzer harmonic oscillator, generating a Mexican-hat type PEC for $H_2$. A proof for quartic PECs is in Section IV. Results are in Section V. Section VI has 6 applications: entanglement, universal function, scaling, probability density function, theory of the chemical bond and the QM $H_2$ PEC. Section VII concludes.

**Known problems with variables/curvatures and shortcomings of parabolic HO**

*(a) Choice of variables and asymptotes for PECs*

Standard PECs use inter-atomic separation r and the scaled Hooke-Dunham variable [18]

$$\delta_D = r/r_0, \; \delta_D - 1 = r/r_0 - 1 \tag{1}$$

However, it is well known that the widely used molecular Dunham function[1,2] $U_{r,D}$, starting off as

$$U_{r,D} = U_0(1-r/r_0)^2 + \ldots = D_e(1-r/r_0)^2 + \ldots = \tfrac{1}{2}k_e r_0^2(1-r/r_0)^2 + \ldots \tag{2}$$

is wrong. It can never converge at large r [19] and needs higher order terms to secure moderate convergence [17], though it uses the correct well depth, covalent bond energy $D_e$ for H+H→$H_2$. The less used inversely scaled Coulomb-Kratzer variable [12]

$$\delta_K = r_0/r, \; \delta_K - 1 = r_0/r - 1 \tag{3}$$

gives an ionic molecular Coulomb-Kratzer oscillator potential

$$U_{r,K} = U_0(1-r_0/r)^2 = D_{ion}(1-r_0/r)^2 \tag{4}$$

This does without higher order terms, is superior in many respects to (2) and always converges to ionic bond energy $D_{ion}$ for $H^+ + H^- \rightarrow H_2$, larger than $D_e$. A choice for $D_e$ or $D_{ion}$ is a choice for (1) or (3). All results of Kratzer model (4) for $H_2$ are given in Appendix A. We remind (4) is also useful for nuclear physics [20].

*(b) RKR turning points as a generic basis of the entanglement of quantum states*

In a PEC, linear turning points $r_\pm$ are at either side of $r_0$; inverse turning points $1/r_\pm$ are at either side of $1/r_0$. To extract these from energies $U_v$, RKR uses f and g, both complex Klein functions of levels $\Delta U_v$, respectively $F(\Delta U_v)$ and $G(\Delta U_v)$ [3], defined as

$$2f = r_R - r_L = F(\Delta U_v); \; 2g = 1/r_L - 1/r_R = G(\Delta U_v); \; f/g = r_R r_L \text{ and } g/f = 1/(r_R r_L) \tag{5}$$

This connection between $\Delta U_v$ and $\Delta r$ leads to continuous PEC $U_r$. However, (5) allows linear as well as inverse turning points at the same time

$$r_\pm = \sqrt{(f/g + f^2)} \pm f = \tfrac{1}{2}(r_+ + r_-) \pm \tfrac{1}{2}(r_+ - r_-) = \tfrac{1}{2}(r_+ + r_-)(1 \pm I) \tag{6}$$

$$1/r_\pm = \sqrt{(g/f + g^2)} \pm g = \tfrac{1}{2}(1/r_+ + 1/r_-) \pm \tfrac{1}{2}(1/r_+ - 1/r_-) = \tfrac{1}{2}(1/r_+ + 1/r_-)(1 \pm I) \tag{7}$$

The reduced difference I between turning points is

$$I = (r_+ - r_-)/(r_+ + r_-) = (1/r_- - 1/r_+)/(1/r_+ - 1/r_-) = (A-B)/(A+B) \tag{8}$$

With (6)-(7), RKR has no preference for $r_\pm/r_0$ in (2) or for $r_0/r_\pm$ in (4). This makes it difficult to understand why inverse Kratzer turning points $r_0/r_\pm$ are hardly used, see (a). Coupling I, $r_\pm$ and $r_0$

$$r_\pm = r_0(1 \pm I)^{\pm 1} \tag{9}$$

conforms to the difference between (1) and (3). As I is independent of a scale factor for $r_\pm$, we use (9) and do not consider generalization $r_{\pm}=r_0(1\pm I)^{\pm n}$, where n is different from 1.

---

[1] Eqn. (2) is part of the full Dunham potential, an expansion in $d_D$ (1), i.e. $U = a_0 d_D^2(1 + a_1 d_D + a_2 d_D^2 + \ldots)$, with Dunham coefficients $a_n$. In a polynomial in (v+½) and J(J+1), i.e. $U = \Sigma Y_{nm}(v+½)^n[J(J+1)]^m$, $Y_{nm}$ are also Dunham coefficients but the complex relation between $a_n$ and $Y_{nm}$ is, however, not relevant here.
[2] A Taylor expansion brings in the very same lead term for the Morse potential [21].

Table 1. $H_2$ data in [5] and this work (r in Å, $U_v$ in eV)($r_0$=0,74173, $D_e$=4,7476, $D_{ion}$=9,7069 [5])

| r [5] | $U_v$ [5] | r reversed[a] | $r/r_0-1$ | $r_0/r-1$ | I | $r=r_0/(1\pm I)$ | $v+½$[b] |
|---|---|---|---|---|---|---|---|
| 0,4109 | 4,729 | 3,2835 | -0,446025 | 0,805135 | 0,777555 | 0,41723 | 14,5 |
| [0,4158][b] | 4,522 | 2,3748 | -0,43945 | 0,783965 | 0,702014 | 0,43579 | 12,5 |
| 0,4319 | 3,88 | 1,8524 | -0,417713 | 0,717365 | 0,621854 | 0,45733 | 9,5 |
| 0,4597 | 2,935 | 1,5148 | -0,380233 | 0,613509 | 0,534363 | 0,48341 | 6,5 |
| 0,5088 | 1,73 | 1,2186 | -0,314036 | 0,457803 | 0,410907 | 0,52571 | 3,5 |
| 0,6337 | 0,269 | 0,8833 | -0,145646 | 0,170475 | 0,164535 | 0,63693 | 0,5 |
| 0,74173 | | 0,74173 | 0,000000 | 0,000000 | 0,000000 | 0,74173 | |
| 0,8833 | 0,269 | 0,6337 | 0,190865 | -0,160274 | -0,164535 | 0,88781 | 0,5 |
| 1,2186 | 1,73 | 0,5088 | 0,642916 | -0,391326 | -0,410907 | 1,25910 | 3,5 |
| 1,5148 | 2,935 | 0,4597 | 1,042253 | -0,510345 | -0,534363 | 1,592937 | 6,,5 |
| 1,8524 | 3,88 | 0,4319 | 1,497405 | -0,599584 | -0,621854 | 1,96149 | 9,5 |
| 2,3748 | 4,522 | [0,4158][b] | 2,201704 | -0,687666 | -0,702014 | 2,48914 | 12,5 |
| 3,2835 | 4,729 | 0,4109 | 3,426813 | -0,774104 | -0,777555 | 3,33444 | 14,5 |
| [4,23[c] | 4,745 | | 4,702884 | -0,824650 | | | ?] |
| [6,35[c] | 4,747 | | 7,561067 | -0,883192 | | | ?] |

[a] r in reversed order absent in [5], since entanglement was not considered
[b] not given in [5] but computed here to have a complete set
[c] these last two energy values in [5] are not observed values, see also Table B1

By definition, (6)-(7) expose state entanglement. They create two sets of algebraically connected turning points for $U_r$: one in increasing order (from small to large), the other in reverse order (from large to small), as shown in Table 1. Using both sets gives an experimentally validated knot or braid view on $H_2$, discussed further below.

*(c) Parabolic behavior of levels $U_v$ in function of v+½*

$H_2$ levels are parabolic in v+½, if zero point energy ZPE is included. Integer v applies for levels without ZPE [13](see Appendix B). These obey the inverse Kratzer variable (5), rewritten as

$$r_0/r_+ - r_0/r_- = 1/(1-½Av\omega/D_{ion}) - 1/(1+½Av\omega/D_{ion}) = Av(\omega/D_{ion})/[1-¼(Av\omega/D_{ion})^2] \quad (10)$$

where ω is the harmonic frequency, constant A is close to 1 (A=0,838 [13]). Ratio $\omega/D_{ion}$= 4410/78844,9=17,87 provides with $v_0=v_{max}$ (see [13] and appendix A). Since $Av(\omega/D_{ion})/[1-¼(Av\omega/D_{ion})^2] \approx Av/v_0+...$, a plot of $U_v$ versus $v/v_0$ gives correct first order curvatures, since $U_v$

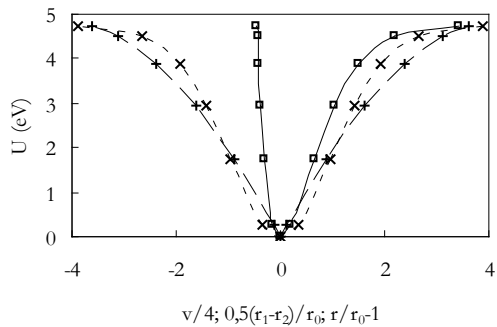

Fig. 1 $U_v$ (eV) versus (v+½)/4 (long dashes +); $r/r_0-1$ (full □) and $½(r_+-r_-)/r_0$ (short dashes x) $\approx[1-(1-v/v_0)^2]+...=2(v/v_0)-(v/v_0)^2+...=2x-x^2+...$ as shown in the right part of Fig. 1 (long dashes

+). This right part can be extended to the left side as in Fig. 1 but on doing so, fitting the complete set in $v/v_0$ becomes problematic. With these 3 problems in mind, we turn to levels and PECs for $H_2$. Fig. 1 shows $U_v$, plotted versus $\pm(v+½)/4$, using scale factor 4 to make this parabola commensurate with the RKR PECs in $r_\pm/r_0-1$ and $½(r_+-r_-)/r_0)$. $U_v$ and $U_{diff}$ approach the asymptote at either side. $U_v$ crosses the origin, which is approached by $U_{diff}$. Although continuous $r/r_0-1$ PEC approaches the origin also, only its attractive (right) branch approaches the limit at large r, due to a curvature switch. Its repulsive (left) branch seems to cross this at small r and no switch in curvature shows. PECs in $r/r_0$ in Fig. 1 cannot be fitted reasonably but PECs in $r_0/r$ in Fig. 2 can. This shows in Fig. 2 with $H_2$ PECs based on inverse turning points. The $U_v$ plot for $\pm v/v_0$ is now compared with continuous $r_0/r_\pm-1$ and difference $½(r_0/r_+-r_0/r_-)$ PECs. A scale factor for $v/v_0$, between -1 and +1, is not needed. Curvatures for PECs are different from those in Fig. 1. PECs are similar and almost left-right symmetrical. The difference PEC obeys quite accurately $2^{nd}$ order (parabolic) and $4^{th}$ order (quartic) fits (see below).

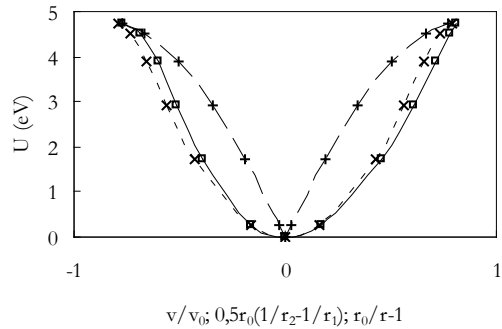

Fig. 2 $U_v$ (eV) plotted versus $v/v_0$ (long dashes +); $r_0/r-1$ (full □) and $½(1/r_+-1/r_-)r_0$ (short dashes x)

The similar curvatures for PECs are opposite to those for $U_v$. Both PECs approach origin as well as limit $U_0=D_e$. These 3 problems (a-c) have important implications.

*(d) Shortcomings of parabolic HO behavior: the confined or closed quartic harmonic oscillator (CHO)*

First and as stated above, PECs in $r/r_0$ cannot be fitted, whereas those in $r_0/r$ can [13,17]. Second, the shape of the $H_2$ PEC depends on the variable: Dunham $r/r_0$ (1) gives asymmetric (Fig. 1), Kratzer $r_0/r$ (3) symmetric PECs (Fig. 2). Since both are compatible with RKR (6)-(7), $r_0/r$ seems superior a variable as to bond symmetry than $r/r_0$, as argued in [17].

Third, the HO parabola with its single extreme, approached by either branch, is deceptive: its infinite branches always cross natural limit $U_0$. With parabolic HO behavior, the $H_2$ dissociation limit would always be crossed or, a HO is always wrong for prototypical natural oscillator $H_2$. While HOs have open branches, natural CHOs have finite branches, confined to $U_0=D_e$. This means that in either branch, a switch of curvatures must occur.

## Confined harmonic oscillator (CHO) for $H_2$: from Kratzer parabola to Kratzer quartic

*(a) Trivial Kratzer HO: turning points in a parabolic PEC*

Without knowing $D_e$, an ionic Kratzer model with difference d between turning points gives

$$U_v = D_{ion}d^2 = \tfrac{1}{4}D_{ion}(r_0/r_- - r_0/r_+)^2 \qquad (11)$$

with trivial solutions

$$d = \pm\sqrt{(U_v/D_{ion})} \qquad (12)$$

Although this parabolic Kratzer HO reproduces levels exactly, its turning points must be tested with those from RKR using (9) with d≈1 ($D_{ion}$=9,707 eV for $r_0$=0,74173).

Fig. 3 shows the $U_v(d)$ plot and RKR PECs in $d_i' = \tfrac{1}{2}r_0(1/r_1 - 1/r_2)$ and $d_i'' = r_0/r_\pm - 1$. Despite the simplicity of (11), RKR PECs are close to this ionic Kratzer PEC. The parabolic Kratzer fit in d

$$U_v = 9,3d^2 \approx D_{ion}d^2 \text{ eV} \qquad (13)$$

has goodness of fit $R^2 \equiv 1$. Ionic Kratzer turning points d are further discussed in Table 3 below.

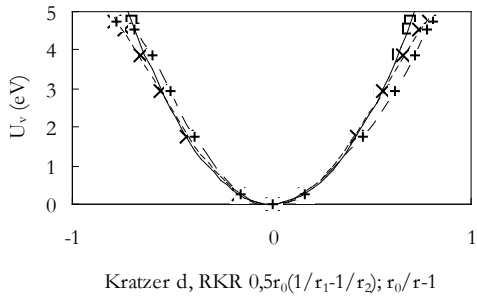 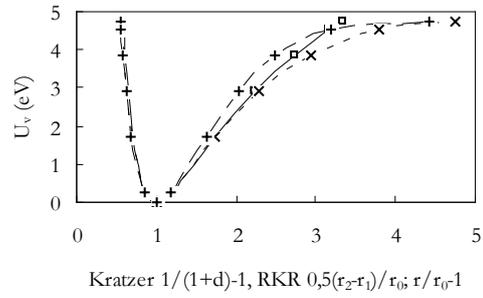

Fig. 3 $U_v$ versus d (full), $d_i'$ (short dashes) and $d_i''$ (long dashes)

Fig. 4 $U_v$ versus $1/(1\pm d)-1$; d' (short dashes) and d'' (long dashes)

In a trivial Kratzer approach, $D_e$ is not required, since any non-zero asymptote $U_0=D$ reproduces levels exactly but turning points must be meaningful. With this criterion, ionic bond energy $D_{ion}$ leads to the best results (see below). Although $D_{ion}$ is much larger than covalent well depth $D_e$, we find that, unlike its depth, the width of the well is governed, for the larger part (see Fig. 4), by ionic Coulomb-Kratzer limit $D_{ion} = \tfrac{1}{2}e^2/r_0$, not by $D_e$ (see Appendix A and [13]).

A quartic fit for the RKR difference PEC using turning points d' [13]

$$U_v = -5,3759d'^4 + 11,089d'^2 \text{ eV} \qquad (14)$$

has a much smaller goodness of fit $R^2 = 0,9982$ (see Table 3 below).

Using (9), the trivial Kratzer approximation is easily extended to variable $r/r_0$. Fig. 4 shows PECs for $1/(1\pm d)-1$ and those for RKR $\tfrac{1}{2}(r_1-r_2)/r_0$ and $r_\pm/r_0-1$. Even with $r/r_0$, a Kratzer potential accounts nicely for 95 % of the complete PEC, which is surprising. While errors for levels are zero (exact result), the difference with RKR turning points is only 4 % (see Table 3). Since trivial

Kratzer HO (13), like all other parabolic HOs, is open and does not behave properly at limit $D_e$, also this plausible Kratzer HO must, mathematically, be upgraded at least to order 4 (CHO).

*(b) From parabola (HO) to quartic (CHO) and Mexican hat potential energy curve*

Observed $U_v$ shows a parabolic dependence on $v/v_0$, see at (10) and Fig. 1-2, with $D_e$ as natural limit (the theory [13] is in (c) below). For any non-zero D, $U_0=D$ and $x=v/v_0$ give a parabola and trivial but always exact solutions, respectively obeying

$$U_v=U_0(2x-x^2)=D(2x-x^2); \quad U_v/D=y=2x-x^2 \text{ and } 1-y=(1-x)^2 \tag{15}$$

$$x_\pm=1\pm\sqrt{(1-U_v/D)}=1\pm\sqrt{(1-y)}=1\pm(1-x) \tag{16}$$

As in (a), (15) returns all levels exactly by definition but parabolic turning points (16) must make sense. An alternative parabola for (15) is obtained with turning points $x'=\pm\sqrt{(2x-x^2)}$ or $y=x'^2$. Again this is still exact for levels but can be meaningless for points x', pending the value of D.

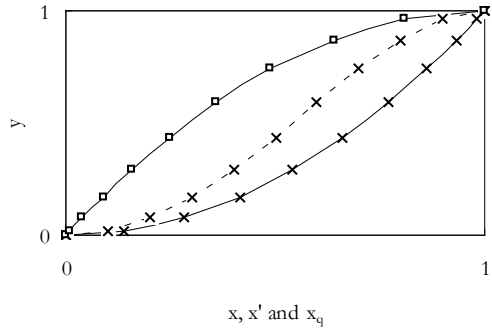

Fig. 5 Plot of y versus x (full □), x' (full x) and $x_q$ (dashes x)

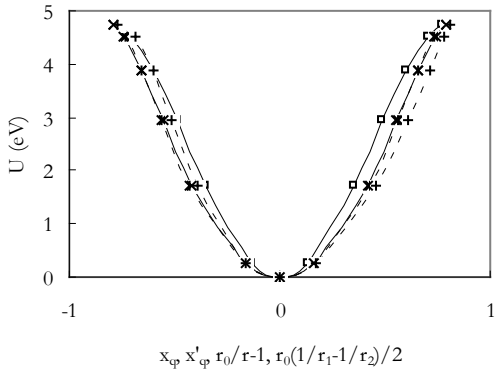 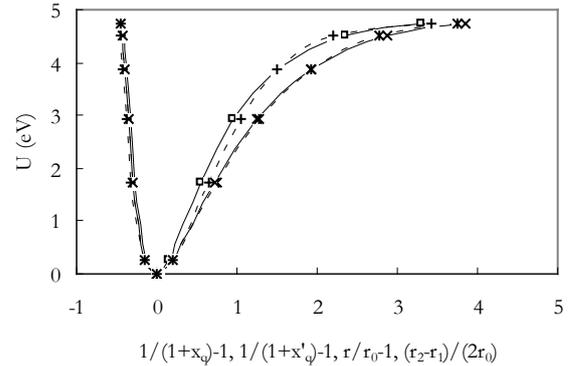

Fig. 6 Kratzer quartic $H_2$ PECs in $x_q$ (full □), x'$_q$ (full x), $r_0/r-1$ (dashes +), ½($r_0/r_1-r_0/r_2$) (dashes *)

Fig. 7 Dunham quartic $H_2$ PECs in $1/(1+x_q)-1$ (full □), $1/(1+x'_q)-1$ (full x), $r/r_0-1$ (dashes +), ½($r_2/r_0-r_1/r_0$) (dashes *)

Mathematically, the simplest way to get a confined HO or CHO is to go over to the square root of x or √x, the generic variable $x_q$ to obtain a Kratzer quartic. With (16), quartic turning points

$$x_q=\pm\sqrt{[1-\sqrt{(1-U_v/D)}]}=\pm\sqrt{x} \tag{17}$$

will also reproduce exactly the very same $H_2$ levels $U_v$ of (15), following the Kratzer-type quartic

$$U_v = U_0(2x-x^2) \equiv U_0[2(\sqrt{x})^2-(\sqrt{x})^4] = U_0(2x_q^2-x_q^4) = U_r \qquad (18)$$

Due to (10), $x_q$ is naturally connected with slightly smaller variable $x'_q$, derived from

$$x_q = x'_q/(1-x'^2_q/4) \text{ or } x'_q = (-2/x_q)[1\pm\sqrt{1+x_q^2}] \qquad (19)$$

The mathematical advantage of quartic (18) is that it creates critical points at either side of the minimum, with accompanying curvature switches (see Fig. 5-6). Given the similarity of $U_v$ and PEC $U_r$ behavior (15)-(18), it is only natural to allow for some scale factor A for $x_q$ to switch from $U_v$ to $U_r$. Kratzer type PECs for A=0,7925≈0,8 in Fig. 6, can be compared with those in Fig. 3; Dunham-type PECs in Fig. 7 with those in Fig. 4. From Fig. 7, it appears that $Ax_q$, is related to $r_0/r-1$ in RKR; $x'_q$ to $\frac{1}{2}r_0(1/r_1-1/r_2)$. Obviously, all 3 Kratzer PECs are of Mexican-hat type, but only the PEC in $0,8x_q$ (X) is exact. Details are in Table 3. The connection of $x'_q$ in (19) with the RKR difference is nearly one by one (see Fig. 7).

*(c) Theoretical treatment using vibrational quantum numbers v+½* [13]

To find a PEC using Kratzer connection (10) between $v/v_0$ and $\Delta U_v$, the values of v+½, not given in [5], are needed. Following the analysis in [13], v+½ is included in Table 1.

A useful application of (15), which is also a stringent accuracy test of a Kratzer bond theory [13], is provided by the transition to complementary variable x'=1-x. This transition obeys

$$U_v = U_0(2x-x^2) = U_0[2(1-x')-(1-x')^2] = U_0(1-x'^2) \qquad (20)$$

which implies that a plot of $U_v$ versus x' gives $U_0$ as *maximum* intercept, when the coefficient for linear x' has vanished since 2x'-2x'=0. With this consistency procedure, we found [13] that

$$x' = a[1/(1-0,022426(v+\tfrac{1}{2}))-1/(1+0,022426(v+\tfrac{1}{2}))] \qquad (21)$$

provides immediately with $D_e$. The result of (21) with the less precise level data in [5] for (20) is 38321,58 cm$^{-1}$ for the intercept and 38259,36 cm$^{-1}$ for the slope. Their average 38290,47 cm$^{-1}$ is close to 4,7476 eV=38292 cm$^{-1}$ in [5]. For the theoretical PEC $U_r$ based on (20)-(21), variable $\sqrt{x'}$ must be used, due to (17).

More precise results would be obtained if more precise observed levels [22] were used as in [13]. We pertain to data in [5], see Table 1, to keep this analysis of the RKR procedure transparent.

**Classical proof for CHOs and quartic PECs instead of HOs**

A classical justification for CHOs is compatible with Klein RKR equations (6)-(8). Using $r_A=r_+$ and $r_B=r_-$ for $r_\pm$ in (9), all possible analytical forms for variables and PECs, centered at the origin, are in Table 2. With both Dunham and Kratzer variables (1) and (3), parabolic HOs and PECs to order 2 are easily completed with 4$^{th}$ order CHOs or PECs, pending the algebraic choices for (9). Since only one set of observed differences $\Delta U_v$ is available, standard solution $\Delta U_v/D_e=I$ gives a difference PEC in 2$^{nd}$ order $I^2$ (see lines for f and g in Table 2). However, interpreting the same difference with continuous PECs in $(r_{AB}/r_0-1)^2$ or $(r_0/r_{AB}-1)^2$, where $r_{AB}=½(r_A+r_B)$, gives $\Delta U_v/D_e = I^2$ and PECs in 4$^{th}$ order $I^4$. This classical proof justifies the CHOs needed. Zero PEC-cases are of interest, since the straight line prediction can immediately reveal if $r_A$ and $r_B$ obey (9).

Table 2. Inventory of all variables and PECs through the origin possible with algebraic (9).

| Variable | Option (a) $r_\pm=r_0(1\pm I)$ | PEC[a] | Option (b) $r_\pm=r_0/(1\pm I)$ | PEC[a] |
|---|---|---|---|---|
| $s=r_{AB}/r_0=½(r_A+r_B)/r_0$ | 1 | $(1-s)^2=0$ | $s=1/(1-I^2)$ | $\mathbf{I^4/(1-I^2)^2}$ |
| $(f/g)/r^2_0=r_Ar_B/r^2_0$ | $1-I^2$ | | $(f/g)/r^2_0=1/(1-I^2)$ | |
| $f=½(r_A-r_B)/r_0$ | $I$ | $f^2=I^2$ | $f=I/(1-I^2)$ | $I^2/(1-I^2)^2$ |
| $s'=½r_0(1/r_A+1/r_B)$ | $1/(1-I^2)$ | $\mathbf{I^4/(1-I^2)^2}$ | $s'=1$ | $(1-s'^2)=0$ |
| $g=½r_0(1/r_B-1/r_A)$ | $I/(1-I^2)$ | $g^2=I^2/(1-I^2)^2$ | $g=I$ | $I^2$ |
| $(g/f)r^2_0=r^2_0/(r_Ar_B)$ | $1/(1-I^2)$ | | $(g/f)r^2_0=1-I^2$ | |
| $s''=r_0/r_{AB}=2r_0/(r_A+r_B)$ | 1 | $(1-s'')^2=0$ | $s''=(1-I^2)$ | $\mathbf{I^4}$ |

[a] PECs, expected classically to be of 4$^{th}$ order, are given in bold

With Table 2, the internal consistency of RKR turning points in [5] is easily verified. Scale invariant I cannot only be calculated from (8) but also from sum and product of turning points. Option (b) gives $I_s=\pm\sqrt{(1-r_0/r_{AB})}$ in the bottom row and $I_p=\pm\sqrt{[1-r^2_0/(r_Ar_B)]}$ in row g. This implies that I in (8) be denoted by $I_0$. If state entanglement were to hold exactly as in (9), $I_0=I_s=I_p$. Since all I-values are easily calculated from RKR turning points in Table 1, a comparison with the trivial spectroscopic I-value, i.e. $0{,}8x_q$ (17) or its theoretical value from (21) is in order. The % differences generated by $I_0$, $I_s$, $I_p$, $r_0/r-1$ and $½_0r_0(1/r_A-1/r_B)$ are all given in Table 3 below. With this criterion, $I_s$ with 3,4 % difference is the better RKR choice for $0{,}8x_q$ (see also below). With the relatively large errors for $U_v$ in Table 3, generated by $I_0$, $I_s$, $I_p$, $r_0/r-1$ and $½_0r_0(1/r_A-1/r_B)$, it is difficult to validate $H_2$ RKR turning points in [5].

Table 2 explains why parabolic oscillator (15) and quartic (18) are completely equivalent for levels but return different turning points (see Fig. 5): parabolic turning points are $\pm x$ (16), quartic turning points are (18) $\pm\sqrt{x}$. This is the sole reason why the HO for natural, prototypical oscillator $H_2$ is a quartic CHO, giving a Mexican hat type PEC, with finite branches, approaching limit $U_0=D_e$ at either side of the minimum without ever crossing it.

Mexican hat curves, generated by a CHO, are, however, complementary to bell or Gauss curves. This brings in a probabilistic interpretation for $H_2$, which will be discussed further below.

**Results and interpretation**

Table 3 compares all results with emphasis on internal consistency, analytical simplicity and accuracy. For transparency, we use $U_v$ in [5] instead of those in [22]. Data are in Tables 1 and B1. PEC 1: the trivial Kratzer method (13) for $D_{ion}=9{,}3$ eV, gives 4,2 % difference for 10 out of 12 RKR turning points, excluding the 2 outermost. Differences $\Delta I$ with RKR $½(r_1-r_2)/r_0$ are 4,2 % for 12 and only 2,32 % for 10 RKR points. This is amazing as levels are reproduced exactly. It explains why a trivial Kratzer approach accounts nicely for >90 % of the $H_2$ PEC, see Fig. 3-4.

Table 3. Accuracy and simplicity of Kratzer HO and 2 CHOs for the $H_2$ PEC, compared with RKR [5]

| Nr | Type | variable x | $U_r$ (eV) | $\Delta U$ cm$^{-1}$ | $\Delta U$% | $\Delta r(\Delta I)$% |
|---|---|---|---|---|---|---|
| | | | | | | |
| PECs from this work | | | | | | |
| 1 | Kratzer[a] | $x=\pm\sqrt{(U_v/D_{ion})}$ (13) | $9,30000x^2$ | 0 | 0 | $4,2^b(2,3)$ |
| 2 | CHO[a,c] | $0,8x_q$ in (18) | $-11,590820x^4+14,836250x^2$ | 0 | 0 | 4,19(3,2) |
| 3 | $CHO_{theo}$ | $0,8\sqrt{x'}$ in (21) | $-11,530711x^4+14,803444x^2$ | 5,4 | 0,04 | 4,24(3,2) |
| PECs from turning points $r_\pm$ in [5] from their difference and from $I_0, I_p$ and $I_s$ | | | | | | |
| 4 | CHO[d] | $r_0/r_\pm - 1$ | $-5,75796x^4+1,30523x^3+11,23558x^2-1,101037x$ | 916,8 | 12,52 | |
| 5 | CHO[d] (8) | $I_0=(r_A-r_B)/(r_A+r_B)$ | $-7,606082x^4 + 12,628366x^2$ | 953,0 | 7,80 | |
| 6 | CHO[d] | $\frac{1}{2}r_0(1/r_A-1/r_B)$ | $-5,375916x^4 + 11,089020x^2$ | 573,5 | 3,99 | |
| 7 | CHO[d] | $I_p=\sqrt{(1-r^2_0/r_A r_B)}$ | $-15,608616x^4+ 17,208229x^2$ | 160,9 | 1,77 | |
| 8 | CHO[d] | $I_s=\sqrt{(1-r_0/r_{AB})}$ | $-11,302032x^4 + 14,775512x^2$ | 617,5 | 5,64 | |
| Morse PEC: (a) $U_v$ from RKR turning points; (b) Morse turning points from $U_v$ | | | | | | |
| 9a | Morse | $U_v$ from RKR r | $D_e\{1-\exp[-2,084(r-r_0)]\}^2$ | 463,9 | 3,68 | |
| 9b | Morse | r from $U_v$ | $D_e\{1-\exp[-2,084(r-r_0)]\}^2$ | 0 | 0 | 3,3(-) |

[a] generic methods for respectively the trivial Kratzer ionic HO model using $U_v/D_{ion}$ and the trivial CHO using $U_v/D_e$
[b] only for this HO, the % differences applies for all turning points, except the 2 outer extremes, see Fig. 3-4
[c] coefficient for $x^2$ is $2D_e/0,8^2=3,215D_e=14,83625$; for $x^4$ this is $D_e/0,8^4=2,4414D_e=11,59082$
[d] all variables calculated with the turning points in [5], collected in Table 1

PEC 2: This second trivial method uses solely levels $U_v$ and $D_e$ as input for turning points (18). Comparing this result with RKR using $0,8x_q$ (17) gives a difference of 4,19% for 12 turning points, including the 2 outer ones and 3,2 % for $I_s$. As for PEC 1, all levels are returned exactly. PECs in $\pm Ax_q$ are of Mexican hat-type (Fig. 6), those in $1/(1\pm Ax_q)-1$ are not (Fig. 7).

PEC 3: The theoretical approach has the advantage over PEC 2 that $D_e$ is also directly available from $U_v$ using (20) [13]. Moreover, PEC 3 is almost indistinguishable from PEC 2. Differences for levels of 5 cm$^{-1}$ (0,04 %) are in line with [13]. Improving its precision is possible [23]. Despite their analytical simplicity, PECs 1-3 of this work are of spectroscopic accuracy as they reproduce levels exactly at the expense of small differences (≤4%) with RKR turning points. Since turning points are not observed but are expectation values, pending the models (RKR, IPA), PECs 2-3 are plausible for $H_2$. PEC 2 is trivial but exact. Fully theoretical PEC 3 is less exact but more complete on the physical implications. PECs 2-3 are of Mexican hat type and derive from a CHO, not from a HO, following the proof above.

PECs 4-8: If RKR were really reliable, it is normal to expect that fitting $U_v$ versus its own turning points should be reasonably accurate. Fits to order 4 should therefore give small, if not zero differences for $U_v$. Surprisingly, RKR PECs 4-8 do not give small errors at all (see Table 3). Errors between 161 and 953 cm$^{-1}$ (0,02 and 0,12 eV) in % are much larger than % errors for r(I) with PECs 1-3. This shows that the quality of PECs 2-3 is better than any RKR PEC.

PEC 8: For comparison and without further comments, we give accuracy details for the well-known Morse potential [21], (unjustly) much more used than Kratzer's [12].

Although details are barely visible in graphs, we nevertheless give 2 more figures. Fig. 8 shows the more symmetric behavior of PECs 2-3 in this work and RKR PECs 5, 7 and 8 using inverse (Kratzer) turning points $r_0/r$ (3). Fig. 9 shows the rather asymmetric, conventional PECs 2-3 in this work, RKR PECs 4 and 8 and Morse PEC 9 using linear (Dunham) turning points $r/r_0$ (1).

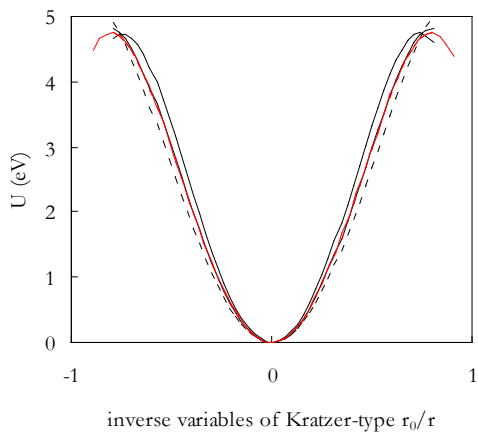 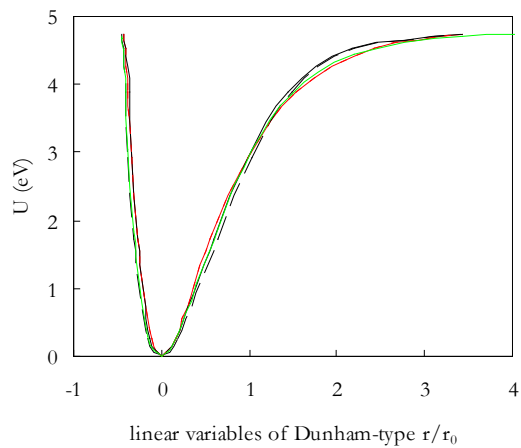

Fig. 8 PEC 2 (red full), 3 (red dashes), 5 (black mixed dashes) 7 (black dashes) and 8 (black full)

Fig 9 PEC 2 (red full), 3 (red dashes), 4 (black long dashes), 8 (black full) and Morse 9 (green)

In Fig. 8, PEC 8 nearly coincides with the single curve for both 2 and 3, as expected from the analytical resemblances in Table 3. We also extrapolated the fit for exact PEC 2 to verify that this is indeed a Mexican hat-type curve, showing that no branch crosses limit $D_e$. However, $H_2$ states beyond the external inflection points are imaginary and cannot be observed according to this theory. In contrast, Fig. 9 suggests that the left branch of all PECs in $r/r_0$ would cross this limit when extrapolated to lower r. This wrong information on prototypical diatomic bond $H_2$ is probably the major shortcoming of PECs in Dunham variable $r/r_0$ (1) and of the Morse PEC. Despite this, PECs in Fig. 9 appear in many textbooks on molecular spectroscopy, whereas equally valid PECs in Fig. 8 with inverse turning points are hardly used.

The rather disappointing accuracy tests for PECs 4-8 are not really surprising, reminding that RKR evolved from a graphical (Rydberg [3a]) to a semi-empirical WKB approximation (Klein-Rees [3b,c]) and that a number of complicated steps is needed to link $\Delta U_v$ with $\Delta r$ (see for instance [24]). Since we succeeded in using only a single and analytically simple step to establish this link between $\Delta U_v$ and $\Delta r$, more applications emerge.

**Applications**

(a) Entanglement of $H_2$ quantum states

Understanding entanglement of spatially separated quantum states is not only important for the EPR-paradox and Bell inequalities [15], quantum information theory and computing [16], topological entropy, non-locality… but also for understanding DNA. We illustrate entanglement of spatially separated $H_2$-states, essentially given away by RKR as shown in Section II. We interpret observed $H_2$ frequency gaps $\Delta U_v$ as radial differences (separations) $\Delta r$ on field axis r.

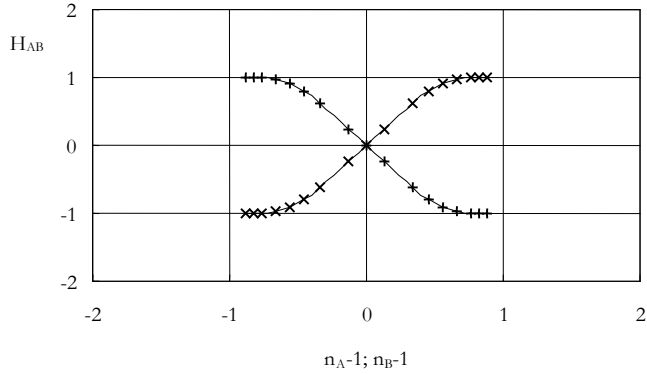

Fig. 10 Plot of $H_{AB}$ versus $n_A-1$ for $H_A$-states (full x) and $n_B-1$ for $H_B$-states (dashes +)

Absolute differences are $|r_A-r_0|$ for atom $H_A$ and $|r_B-r_0|$ for atom $H_B$ if $r_0$ is the equilibrium bond length. With $H_A$ far away at $r_A=r_0/(1-I)\gg r_0$ (I=0,99 gives $r_A=100r_0$), its dimensionless number is

$$n_A=r_0/r_A=1-I \text{ or } n_A-1=-I \qquad (22)$$

Classical RKR physics (6)-(8) automatically fixes $r_B$ for partner $H_B$ at $r_B=r_0/(1+I)$ or at number

$$n_B=r_0/r_B=1+I \text{ or } n_B-1=I \qquad (23)$$

( I=0,99 gives $0,502r_0$ with lower limit $½r_0=0,37086$ Å at I=1).

Next, the square root of reduced $U_v/D_e$ gives similar numbers for the spectrum of bond $H_2$, i.e.

$$H_\pm=H_{AB}=\pm\sqrt{(U_v/D_e)} \qquad (24)$$

A combination of all these numbers for $H_2$ leads to a braid plot of $H_{AB}$ versus $n_A-1$ and $n_B-1$ as shown in Fig. 10. The advantage of a braid view over that with PECs/CHOs, as above, is that it nicely illustrates why the choice for $H_A$ and $H_B$ is arbitrary. Interchanging suffixes, i.e. allowing for $H_B$ and $H_A$ also, is allowed and even essential. This is also a cornerstone of QM theory of the chemical bond (LCAO, VB and MO methods [2]). Entanglement of H-states at any separation from the center, large or small, is clearly exposed in the braid view in Fig. 10 and by (22)-(23). Conceptually, entanglement of $H_2$ states is secured by the fact that the very same frequency gaps $\Delta U_v$ are used to fix radial separations $\Delta r$ at either side of the minimum. Braid views are important for (anyon) theory, following [25] in a comment on recent experiments on entanglement [26].

(b) Universal molecular potential energy function (UPEF)

The importance of a universal function is evident when looking at the 5 PECs in Fig. 11 for $H_2$, HF, $I_2$, $N_2$ and $O_2$ in [5]. Scaling $U_v$ and r or 1/r proceeds with (i) numbers $U_v/D_e$ for the well depth and (ii) either Dunham $r/r_0$ (1) or Kratzer numbers $r_0/r$ (3) or combinations like I (9). Fig. 12 shows that these scaling effects do not lead to a unification of the 5 PECs in the $r/r_0$-mode.

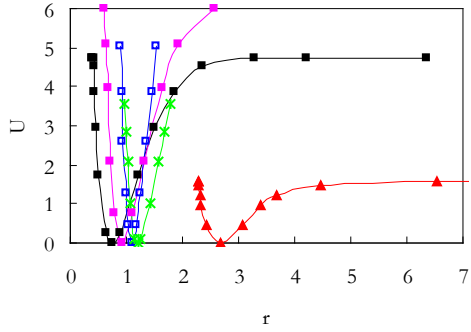
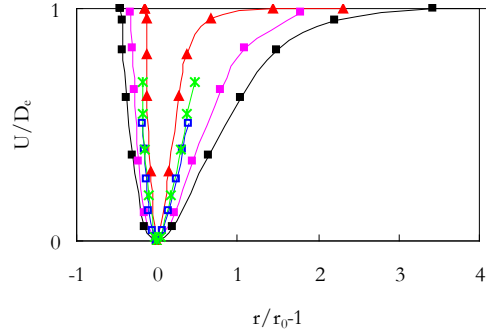

Fig. 11 RKR PECs for $H_2$ (black), HF(pink), $I_2$(red) $N_2$(blue) and $O_2$(green)

Fig. 12 Reduced PECs in $r/r_0$, centered at the origin (same colors)

For Kratzer representations in $r_0/r$, centered at the origin also, we choose $I_s$ in Table 2-3. This gives Fig. 13. The asymmetry of PECs in Fig. 11 and 12 may be removed but, exactly as in Fig. 12, a single (universal) PEC is not yet obtained.

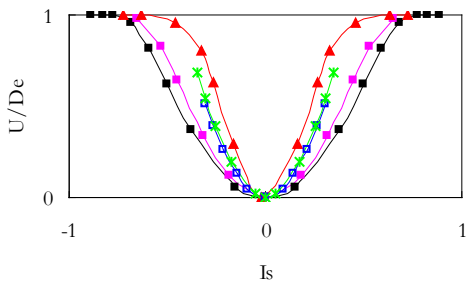
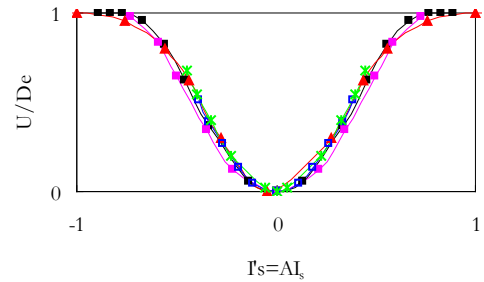

Fig. 13 Reduced PECs versus Kratzer-type $I_s$

Fig. 14 Reduced PECs versus scaled $I'_s$

To make all PECs coincide, the essential role of a UPEF, an extra scale factor is needed for $I_s$, as $U_v$ is properly normalized with $D_e$. Since the width of the well is determined by $r_0$ or by $D_{ion}$, as discussed above, this extra scale factor A should vary with $r_0$. We found (empirically) that

$$A=(r_0/r_{H2})^{1/4} \text{ and } I'_s=AI_s \tag{25}$$

led to Fig. 14, which must be considered as a first attempt to produce a UPEC. Fig. 15 gives corresponding asymmetric PECs in $1/(1\pm AI_s)-1$, where PECs only nearly coincide. Unification is not yet perfect but the results strongly favor the existence of a UPEF. In reality, see Introduction, the genuine UPEC is already available by applying the results for $H_2$ above directly to all bonds. Plotting the 5 $U_v/D_e$ sets versus $0,8x_q$ in (17)-(18) gives the genuine UPEC, shown in Fig. 16.

Unlike Fig. 15, Fig. 16 clearly shows that a single asymmetric PEC is generated for all 5 bonds with $1/(1\pm0,8x_q)-1$. The perfectly symmetric PEC of Mexican hat-type with $\pm0,8x_q$ is in Fig. 17. In Fig. 16, the full line is an aid to the eye, since a fit for these asymmetric PECs is impossible.

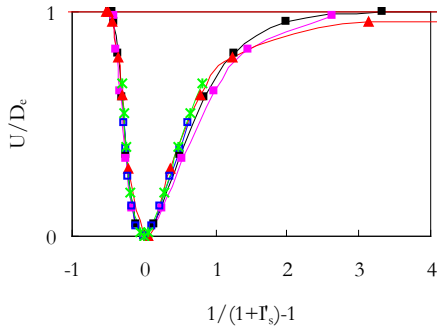
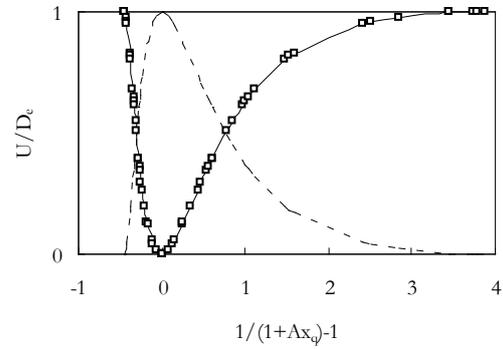

Fig. 15 Nearly coinciding PECs due to scaling by $r_0$

Fig. 16 Trivial, exact asymmetric UPEC (full), complementary UPEc (dashes)

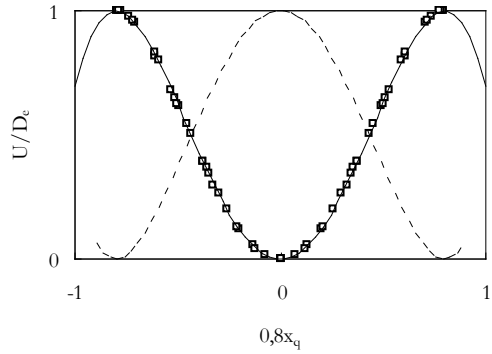

Fig. 17 Trivial symmetric UPEC of Mexican hat type (full line),
complementary symmetric UPEC of Gauss type (dashes)

Fig. 16 also shows the complementary UPEC for $1-U_v/D_e$, which is, by definition, as asymmetric as the normal UPEC. The situation is different for the symmetric UPEC in Fig. 17. Here, the curve is an exact 4$^{th}$ order fit, given in Table 3. Extrapolated as in Fig. 8 for H$_2$, this symmetric UPEC is indeed a Mexican hat curve. The fit for the complementary $1-U_v/D_e$ (dashes) is equally exact, since trivial, but, in addition, this gives away a bell curve, typical for a normal Gauss distribution, which suggests that exponential fitting may be successful too (see further below). The covalent UPEC, based on a universal CHO, must not detract us from the fact that the Kratzer HO and corresponding Kratzer PEC for H$_2$ is fairly accurate, as shown in Section III (a). We cannot be complete, unless the trivial ionic UPEC of HO-type is also considered, for which numbers $U/D_{ion}$ must be used, instead of $U/D_e$. Both universal but trivial HO and CHO's are compared in $r/r_0$ and $r_0/r$ version. First, we plot $U_v/D_e$ versus $x_q$ and $U_v/D_{ion}$ versus d (12). For the 5 bonds, the ionic and covalent UPECs in the $r_0/r$-view are in Fig. 18.

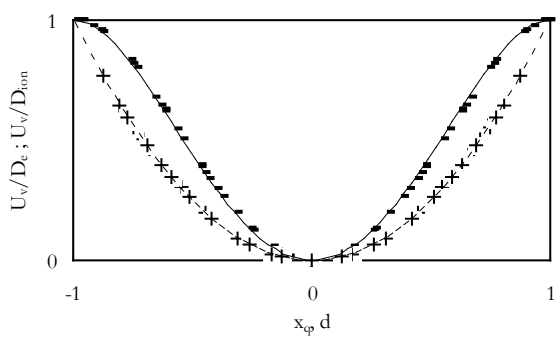
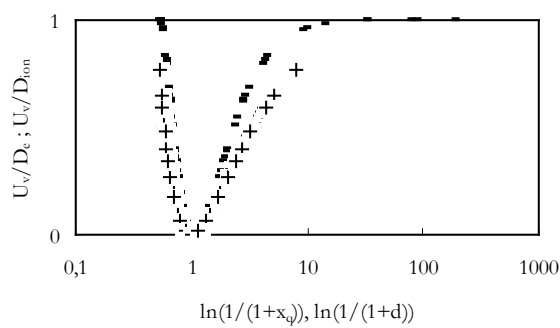

Fig. 18 $U_v/D_e$ vs $x_q$ (full -), $U_v/D_{ion}$ vs d (dashes +):$r_0/r$-mode

Fig. 19 Same as Fig. 18 but in $r/r_0$-mode

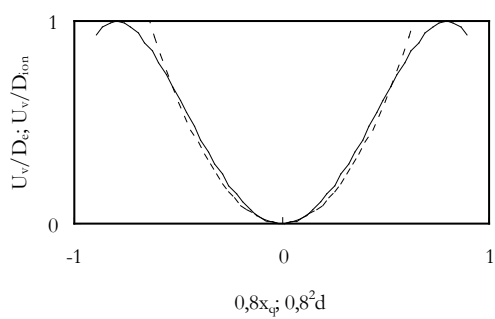
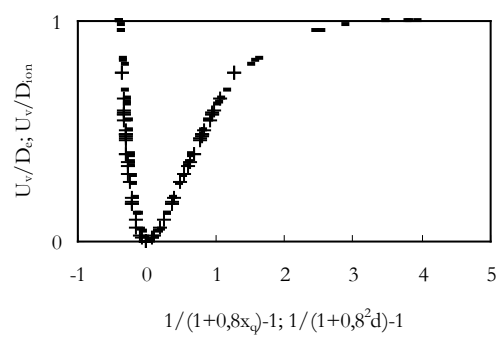

Fig. 20 $U_v/D_e$ vs $0,8x_q$ (full), $U_v/D_{ion}$ vs $0,8^2d$ (dashes):$r_0/r$-mode

Fig. 21 Same as Fig. 20 but in $r/r_0$-mode

For the $r/r_0$-mode in Fig. 19, the largest $r/r_0$ value ($H_2$ level $0,99988D_e$) is $1/(1-0,994363)=177,4$. This enforces a logarithmic scale for the x-axis. Scaling by $D_e$ leads to extremely large r-vales for the right branch, which are avoided with scaling by $D_{ion}$. Moderating $x_q$ by factor 0,8, leads to $1/(1\pm 0,8x_q)$, much more in line with RKR but it may be questioned whether or not this moderation is realistic. For the UPEC, this factor enlarges the 4th order term by $(5/4)^4=2,441$ (see Table 3). Adapting the ionic UPEC similarly, leads to the situation in Fig. 20-21, where the two PECs are much closer together and whereby very large r-values in Fig. 19 are avoided.

Fig. 20-21 illustrate how difficult it is to distinguish HO- from CHO-behavior. They equally illustrate how difficult it is to distinguish 19th century based ionic from modern (QM) covalent chemical bonding in the neighborhood of the minimum (see further below on ionic models). This leads us to a broader view on the role of $D_{ion}$ as scaling aid for molecular constants.

(c) Comparing $D_{ion}$ and $D_e$ as scaling aids for 300 low order molecular constants

Since (25) suggests that the ultimate extra scaling factor to expose the reality of a UPEF depends on $r_0$ or $D_{ion}=\frac{1}{2}e_2/r_0$, we wished to verify this scaling hypothesis on a much larger scale. We compare scaling efficiencies of $D_{ion}$ and $D_e$ for over 300 bonds [7] using vibrational constant $\omega_e x_e$, i.e. the coefficient $Y_{20}$ for $(v+\frac{1}{2})^2$ in the Dunham expansion[1]. The 314 bonds for which $\omega_e x_e$ was known in 1999 [7] are reviewed here without any prior selection: 95 bonds contain H, D or T with maximum $\omega_e x_e$=121,336 cm$^{-1}$ for $H_2$ [27]; 219 other bonds have maximum 16,262 cm$^{-1}$ for $NO^+$ [27]. For $D_e$ and $D_{ion}$, scaling maxima are 90000 cm$^{-1}$ (≈89462 cm$^{-1}$ for CO [27]) and 80000 cm$^{-1}$ (≈78321 cm$^{-1}$ for $H_2$ [27]).

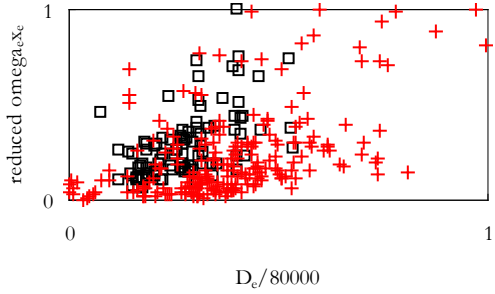
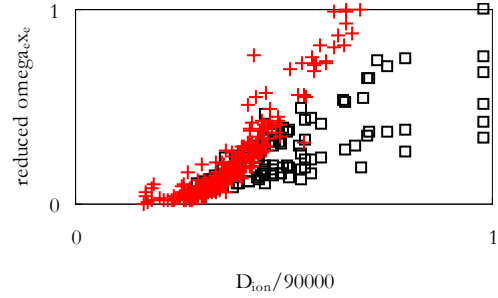

Fig. 22 Plot of reduced $\omega_e x_e$ vs reduced $D_e$ (□ with H,D or T, + without)

Fig. 23 Plot of reduced $\omega_e x_e$ vs reduced $D_{ion}$ (□ with H,D,T, + without)

Fig. 22-23 reveal that, while there is almost no order in the plot with $D_e$, points are well ordered with $D_{ion}$. This confirms our conclusion above that scaling $\omega_e x_e$ will be more successful with $D_{ion}$ than with $D_e$. Moreover, the spread of points for H,D,T-bonds in Fig. 23 closely follows a reduced mass classification: in the 6 classes for H, D, T (marks □), the upper series is for H-, the lower for T-containing bonds. This proves that for the spectroscopic constants, the scaling power of $D_{ion}$ or $r_0$, as in (25), is greater that of $D_e$, as argued before [7].

More examples to illustrate this superior scaling power of $D_{ion}$ for lower order constants (Dunham coefficients) are available [6-7,13-14] and must not be repeated here. Theoretically, this scaling power is now understood with the width ($D_{ion}$ or $r_0$), not the depth of the well ($D_e$).

(d) Quantum information theory, Gauss probability density function for diatomic bonds

For 5 bonds, the universal Mexican hat curve and its complementary bell-curve are shown in Fig. 17. Table 2 shows that variable $(1/(1-I^2)$ will give a quartic PEC of type $(1/(1-I^2)-1)^2=I^4/(1-I^2)^2$ but, thus far, this functional was not yet detected.

However, we analyzed the bell-curve in Fig. 17, starting from the standard Gauss distribution of differences x from a mean value, given by

$$G(x)=e^{(-\frac{1}{2}x^2)}/\sqrt{(2\pi)} \quad (26)$$

where mean μ is zero and standard deviation $\sigma^2$=1. To apply (26) for $H_2$ and the other bonds in Fig. 11-17, we used PEC variable $(1/(1-I^2)-1)$ in Table 2 and reformed (26) in

$$1-U_v/D_e=G_I=e^{\{-A[1/(1-I^2)-1]\}} \quad (27)$$

where A is a constant, related to $r_0$, and where I=0,8$x_q$, as above. With A=2,945, linear fit

$$(1-U_v/D_e)= 1{,}00685e^{\{-2{,}945[1/(1-I^2)-1]\}} \quad \text{or} \quad U_v \approx D_e(1-G_I) \quad (28)$$

has a goodness of fit $R^2=0{,}99996$ and gives an average error for all 5 bonds of 140 cm$^{-1}$. Although (27)-(28) give a phenomenological bonding approximation on the basis of a universal Gauss probability density function (PDF), (28) is more precise than any of the RKR PECs for H$_2$ (see Table 3). We remind that exponential fitting (28) always remains less accurate than fitting to order 4 in $x_q$, which is always exact due to (17), even for complementary $1-U_v/D_e$. This is evident from (18) and (20). Since $U_v/D_e=2x_q^2-x_q^4$ is, by virtue of (17), exact by definition, complementary $1-U_v/D_e=(1-x_q^2)^2$ is as exact. The curve can be approximated by a Gaussian like (26) but this is always accompanied with a loss of precision. These questions were already asked at the earlier stages of QM [15].

In practice and in terms of probabilities, intimately connected with wavefunctions in QM, (28) also confirms Kohn-Sham density functional theory models [28]. It shows the effect of Gauss type orbitals (GTO) and Slater type orbitals (STO) for the description of bonds, especially H$_2$. Extrapolating the quartic, obtained by plotting Gauss function (28) versus $0{,}8x_q$, reveals that also this curve has critical points at $\pm 0{,}8x_q$, as illustrated in Fig. 17.

In a less accurate bell curve approach, statistics and probabilities behind the universal PDF (UPDF) would result from a Bernoulli trial with a diatomic bond, confined to an infinitesimally thin coin. If heads were H$_{AB}$, referring to H$_A$+H$_B$→H$_2$, tails are H$_{BA}$, referring to H$_B$+H$_A$→H$_2$. At the top of the bell curve for H$_2$, uncertainty (certainty) about its origin is maximum (minimum). Only at the bottom of the bell curve, it is certain from which combination H$_{AB}$ or H$_{BA}$, H$_2$ was formed. The H$_2$ coin is perfect, since H$_{AB}$ and H$_{BA}$ have equal probabilities for this homonuclear bond. However, a coin like model also implies that H$_{AB}$ and H$_{BA}$ exhibit chiral symmetry. This is easily understood, if the thin H$_2$ coin were a transparent sheet with marks + and – embedded. Tossing would result in marks in order +, - (heads) or in order -,+ (tails). Chiral symmetry for H$_2$ is in line with Hund [29], who showed that chiral symmetry is given away with a Mexican hat curve. Such curve was even detected in the Lyman series of atom H [30].

Nearly accurate probability/uncertainty approximations for H$_2$ may be explained by Heisenberg's principle [31] but they are obviously also connected with classical entropy (disorder) and with binary entropy as defined in information theory (Shannon [32], Fisher [33], R'enyi [34]).

(e) Theory of the chemical bond

A 5$^{th}$ application regards the 10-term QM Hamiltonian **H** for prototypical H$_2$. Using the standard notation for 4 charged particles: 2 leptons a, b and 2 nucleons A, B, we have

$$U=\tfrac{1}{2}m_a v_a^2+\tfrac{1}{2}m_b v_b^2+\tfrac{1}{2}m_A v_A^2+\tfrac{1}{2}m_B v_B^2-e^2/r_{aA}-e^2/r_{bB}\pm(-e^2/r_{bA}-e^2/r_{aB}+e^2/r_{ab}+e^2/r_{AB}) \quad (29)$$

with 4 kinetic energy terms and 6 Coulomb potential energy terms. Reducing the 10 terms in

$$\mathbf{H} = \frac{1}{2}\Sigma(p_x^2/m_x) - e^2\Sigma(1/r_x) \tag{30}$$

is possible by symmetry considerations [17] but a classical alternative exists. The information on the uncertainties for the $H_2$ ground state, as described with a Gauss PDF and obeying a Bernoulli trial for coin-like $H_2$, derives from an ionic Kratzer HO or CHO model. The original 19[th] century ionic model is of electrostatic DC-type with anion and cation preserved, without dynamics but with a built in permanent electric dipole moment. The Gaussian view on $H_2$ learns that this ionic approach must be of electrodynamic AC-type, whereby it is uncertain which of the 2 bonding partners is the anion or the cation and whereby a permanent electric dipole moment is no longer needed. Still, ionic AC type bonding fully complies with the Kratzer bound state equation A(1)-A(2). For (29)-(30), this ionic AC model proves extremely useful. It leads to only one kinetic energy term instead of 4 and to only one Coulomb term instead of 6. For 2 atoms H, reduced mass is $\mu = m_H^2/(2m_H) = \frac{1}{2}m_H$. Two ions lead to a pseudo one-particle system with reduced mass

$$\mu_i = (m_H + m_e)(m_H - m_e)/(2m_H) = \frac{1}{2}m_H(1 - m_e^2/m_H^2) = \frac{1}{2}m_H(1 - 2{,}963^{-7}) \approx \frac{1}{2}m_H = \mu \tag{31}$$

Using levels U precise to order 0,001 eV, this small difference can safely be neglected.

With Coulomb attraction $-e^2/r$ between 2 ions in whatever order, we are left with only 2 terms

$$U_i = U_{ionic}(r) \approx \frac{1}{2}\mu_i v^2 - e^2/r \text{ and } \mathbf{H}_i \approx \frac{1}{2}p_i^2/\mu_i - e^2/r \tag{32}$$

a considerable, if not drastic simplification over the 10 terms in (29) and (30) [13].

Next, for the equilibrium situation at $r_0$, we can use A(5) or

$$\mu_i v_0^2 = e^2/r_0 \tag{33}$$

Plugging this in (32) returns

$$U_{i,0} = -\frac{1}{2}\mu_i v_0^2 = -\frac{1}{2}e^2/r_0 \tag{34}$$

Since (32) is compatible with the original ionic Kratzer potential A(1)-A(2), $U_i$ varies as

$$U_i = U_0(1 - (1 - r_0/r)^2 = 2r_0/r - (r_0/r)^2 = 2x - x^2 \tag{35}$$

formally identical to (15), the basis of the present analysis. Ionic bound state equation (35) is much simpler than (29)-(30), even for prototypical covalent bond $H_2$, and with the results for $H_2$ in Appendix 1, the ab initio character of this approach is well established.

Furthermore, (34)-(35) associate vibrations with the heavier particles, which is also the basis of the Franck-Condon recipe [35]. This led to Franck-Condon factors (FDFs), important for the interaction between ground and excited states. To expose more symmetry details, it may be useful to look at the vibrational spectra of excited states in a $r_0/r$ mode too, rather than in a conventional $r/r_0$ mode, which may help to calculate precise FDFs [36].

The numerical bell curve for dimensionless $H_2$ as described by $U_v/D_e$, is easily associated with the exponential in the Morse potential and, more generally, with wavefunctions. Rather than giving analytical details on these connections, however interesting, we compare our $H_2$ PEC directly with the $H_2$ PEC from ab initio QM.

(f) $H_2$ PEC by ab initio QM [37]

The $H_2$ PEC being so crucial for theory, it aroused the interest of theorists for nearly a century [37-43], starting with Heitler and London [43] in 1927. Ab initio QM calculations are tedious. Many parameters are needed and hundreds of wavefunctions are used [37]. Unfortunately, and as remarked in [11,23], none of the many QM PECs was ever published in analytical form: instead, U and r are tabulated. For the 1993 $H_2$ QM PEC, Wolniewicz [37] lists 54 r-values between 0,2 and $12r_B$, where $r_B$ is the Bohr length, for which U is calculated. Points in this QM PEC are so close to the RKR PEC that a distinction between the 2 in a graph is impossible. As above (see also Table B1), we restrict the analysis to the 14 observed $H_2$ levels [22]. This leaves 33 points between $0,8r_B$ and $6r_B$. The absolutely exact, since trivial, PEC in $x_q$, as defined in our work by (17)-(18), i.e.

$$U_v = 2D_e x_q^2 - D_e x_q^4 = -38292,984 x_q^4 + 76585,968 x_q^2 \text{ cm}^{-1} \quad (36)$$

reproduces 33 $U_v$ exactly. From r in [37], we calculated $r_0/r-1$ to find a link with $x_q$ in (36). The difference between $0,8x_q$ in this work and $r_0/r-1$ from [37] is only 2,6 % for 33 levels. This is better than for the RKR PECs in Table 3. The QM $H_2$ PEC, graphically indistinguishable from the RKR $H_2$ PEC [5], and the PEC in this work are given in Fig. 24, where the Morse PEC is included for reference.

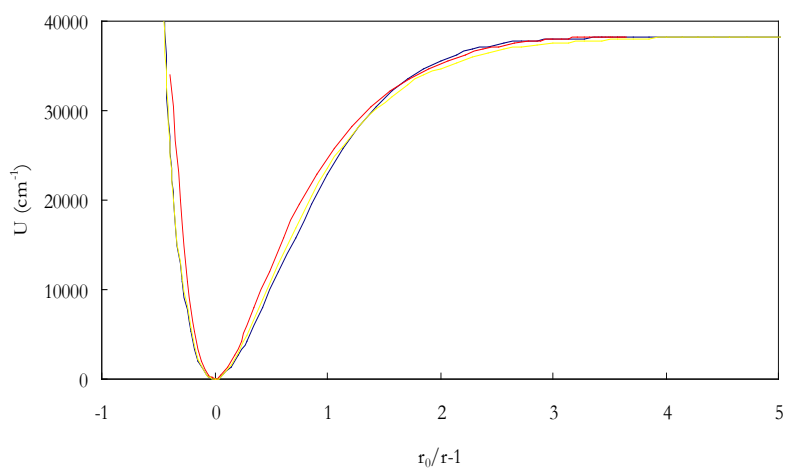

Fig. 24 Four $H_2$ PECs according to QM and RKR (black), this work (red) and Morse (yellow).

The general shape of the $H_2$ PEC is respected in all 4 schemes but discrepancies remain on curvatures between minimum and limit. However, of all 4 methods, only ours reproduces all $U_v$ exactly with a simple analytical function (36), not provided by QM. In [37], $U_v$ for $r=0,2r_B$ is greater than 700000 cm$^{-1}$ or 18 times limit $D_e$. This contradicts the essence of this work, stating that either branch of the $H_2$ PEC is limited to $D_e$ and that the lower limit for r in the left branch is $½r_0$ (0,7$r_B$ or 0,41 Å). With this extremely large $U_v$-value for r=0,2$r_B$, ab initio QM seems to have been inspired by the empirical Morse potential, which tolerates crossing of $D_e$ at the repulsive branch, see Fig. 24. Similarly, the limit for the right branch is 7$r_B$ or 3,7 Å, if A in $Ax_q$ is constant and equal to 0,8=4/5. It remains to be verified whether or not this is acceptable.

**Conclusion**

A quartic Kratzer confined harmonic oscillator (CHO) for ground state $H_2$, itself hidden in the original Kratzer HO, implies that the theory of the chemical bond is much simpler than believed. This CHO gives rise to a $H_2$ PEC of Mexican hat-type, consistent with an ionic model, even for $H_2$ [45]. The long-standing problem about the universal molecular function (UPEF) is solved satisfactorily and analytically. Unlike QM, accurate PEFs and PECs are not only simple; they are analytically available. Entanglement of long- and short-range hydrogen quantum states is a classical, elementary arithmetic, physics element of RKR, itself a WKB approximation. Complementary PECs can be interpreted as a Gaussian but only with loss of precision, which may be important for the EPR-paradox and related problems. Accepting small discrepancies, quantum information theoretical concepts (entropy) seem to appear in a natural way. If so, $H_2$ would not only be prototypical for classical and modern physics and chemistry, for HO/CHO, for braid (anyon) theory and quantum information theory but even for bonding in DNA [X]. At the Bohr scale, a Kratzer-Coulomb law smoothly accounts for minimum and extremes at either side of this minimum for bond $H_2$, the prototypical 4 unit-charge system with 4 charges pair-wise conjugated. Since we dealt successfully with 2-nucleon system $H_2$ in the context of a RKR-WKB model, a similar Kratzer scheme may be useful at the nuclear scale [20], where nucleon interactions are the rule.

**Acknowledgements.** I greatly appreciate my contacts with Y.P. Varshni (U Ottawa).

Appendix A. Kratzer predictions for bond $H_2$ [13]

Conceptually, the original *ionic* Kratzer potential [12]

$$U_K = -e^2/r + B/r^2 \qquad A(1)$$

behind (4) in the text has advantages [13]. First, searching for a minimum with $dU/dr=0$ provides with $B = \frac{1}{2}e^2 r_0$ and gives

$$U_K = -e^2/r + \frac{1}{2}e^2 r_0/r^2 = \frac{1}{2}(e^2/r_0)[2r_0/r - (r_0/r)^2] = D_{ion}(2x - x^2) = U_0(2x - x^2) \qquad A(2)$$

Secondly, searching for the force constant with $d^2U/d^2r = k_e$ leads to

$$k_e = e^2/r_0^3 \qquad A(3)$$

This analytical result for $k_e$ is impossible with Dunham's potential, which needs observed U-data to get $k_e$. Dunham's procedure $\frac{1}{2}k_e r_0^2 = a_0$ can now be completed with $\frac{1}{2}k_e r_0^2 = a_0 = D_{ion} = \frac{1}{2}e^2/r_0$. As a result, $H_2$ fundamental frequency $\omega$ is obtained directly from A(3) since

$$2\pi\omega = \sqrt{(k_e/\mu)} = \sqrt{[2e^2/(m_H r_0^3)]} \qquad A(4)$$

This equilibrium result can also be written as

$$\mu v_0^2 = e^2/r_0 \qquad A(5)$$

Finally, if $r_0$ were available from mass $m_H$, the complete Kratzer $H_2$ HO and parabolic PEC are available without any other assumption [13]. To reach a full *ab initio* solution for this HO, the classical 19[th] century relation between mass, density $\gamma$ and radius of sphere-like H [13] gives

$$m_H = (4\pi\gamma/3)r_0^3 \qquad A(6)$$

Assuming only $m_H$ is known ($m_H = 1837,15267247 m_e$ [44]) and that $\gamma = 1$, A(6) leads to

$$r_0 = 0,736515 \text{ Å} \qquad A(7)$$

Table A1 shows all $H_2$ spectroscopic characteristics derived from Kratzer potential A(1).

Table A1 $H_2$ constants from Kratzer potential and the $H_2$ spectrum [22,27] ($m_H = 1,67353 \cdot 10^{-24}$ g [46])

|  | Kratzer result | from the spectrum [18] | % diff |
|---|---|---|---|
| ($\mu = \frac{1}{2}m_H$ | $8,367663 \cdot 10^{-25}$ g | also assumed known) | |
| $r_0$ | 0,73652 Å | 0,74144 Å | -0,66 |
| $\omega$ | 4410,173 cm$^{-1}$ | 4401,213 cm$^{-1}$ | 0,20 |
| $k_e$ | 577452,8805 force/cm | 575108,9041 force/cm | 0,41 |
| $a_0 = D_{ion}$ | 78844,9005 cm$^{-1}$ | 79578,4482 cm$^{-1}$ | -0,92 |
| $D_0$ (a) | 36110,245 cm$^{-1}$ | 36118,344 cm$^{-1}$ | -0,02 |
| $D_e$ (a) | 38281,14 cm$^{-1}$ | 38292,984 cm$^{-1}$ | -0,02 |

(a) from a recent analysis of the $H_2$ spectrum with a Kratzer variable [23]

The Kratzer $H_2$ HO or the parabolic PEC is completely known with $m_H$ [46] and $D_{ion}$, $k_e$ and $\omega$ directly available from the above equations.

Appendix B Comparing observed $H_2$ levels [22] with RKR levels [5]

Although we adhered to $H_2$ level data as given in [5] for reasons of transparency, more precise data are available [22]. In Table B1, columns 1-3 give $H_2$ data for v, $\Delta G_v$ and $U_v$ as observed by Dabrowski [20]. Columns 4-5 give v+½, not in [5], and levels, adapted for zero point energy ½ω. Columns 6 contains levels in [5], converted with 1eV=8065,479 cm$^{-1}$. To verify the consistency of levels in [5], we calculated ZPE by taking the difference $U_v$ [22]-$U_{v+½b}$[5] in column 6. This shows that, with the levels in [5], ZPE varies between 2166 and 2171 cm$^{-1}$, in line with an expected uncertainty of 8 cm$^{-1}$ for data, given to order 0,001 eV. These discrepancies are too small to have an effect on the theory as applied in this work. However, the levels in the 2 lower rows for long-range behavior of $H_2$ are clearly not observed in [22]. They are not repeated at small r in the original paper either [5], as already indicated in Table 1.

Table B1 Observed $H_2$ levels [22], RKR levels [5] and zero point energies ZPE

| v | $\Delta G_v$ [22] | $U_v$ [22] | v+½ | $U_{v+½}$ a | $U_{v+½}$[5] | ZPE=$U_{v+½b}$[5]-$U_v$[22] |
|---|---|---|---|---|---|---|
| 0 | (4401,21$_3$) | 0 | 0,5 | 2170,85 | 2170 | 2171 |
| 1 | 4161,14 | 4161,14 | 1,5 | 6331,99 | | |
| 2 | 3925,79 | 8086,93 | 2,5 | 10257,78 | | |
| 3 | 3695,43 | 11782,36 | 3,5 | 13953,21 | 13953 | 2171 |
| 4 | 3467,95 | 15250,31 | 4,5 | 17421,16 | | |
| 5 | 3241,61 | 18491,92 | 5,5 | 20662,77 | | |
| 6 | 3013,86 | 21505,78 | 6,5 | 23676,63 | 23672 | 2166 |
| 7 | 2782,13 | 24287,91 | 7,5 | 26458,76 | | |
| 8 | 2543,25 | 26831,16 | 8,5 | 29002,01 | | |
| 9 | 2292,93 | 29124,09 | 9,5 | 31294,94 | 31294 | 2170 |
| 10 | 2026,38 | 31150,47 | 10,5 | 33321,32 | | |
| 11 | 1736,66 | 32887,13 | 11,5 | 35057,98 | | |
| 12 | 1415,07 | 34302,20 | 12,5 | 36473,05 | 36472 | 2170 |
| 13 | 1049,16 | 35351,36 | 13,5 | 37522,21 | | |
| 14 | 622,02 | 35973,38 | 14,5 | 38144,23 | 38142 | 2169 |
| ? | | | | level not measured | 38271[b] | |
| ? | | | | level not measured | 38287[b] | |

a zpe (zero-point energy) 2170,849 cm$^{-1}$ [23]
b Given $\Delta G_v$ for v=14 in Column 2, the last observed level is more than 600 cm$^{-1}$ below the dissociation limit. This implies that these 2 levels in [5] are not observed. These levels and their turning points, critical for long range $H_2$ behavior, may well have been inspired by the Morse-potential [21]. This also justifies the limit of 33, set to the 55 data points for the QM H2 PEC [37] (see text).